\date{10 Jan 2003}

\documentstyle[multicol,pra,aps,amsfonts,epsf]{revtex}

\newcommand{\beq}{\begin{equation}}
\newcommand{\eeq}{\end{equation}}
\newcommand{\beqy}{\begin{eqnarray}}
\newcommand{\eeqy}{\end{eqnarray}}

\newtheorem{Definition}{Definition}
\newtheorem{Lemma}{Lemma}
\newtheorem{Theorem}{Theorem}

\newenvironment{Definition*}{{\bf Definition}}{}

\def\C{{\mathbb{C}}}
\def\Z{{\mathbb{Z}}}
\def\R{{\mathbb{R}}}

\newcommand{\cH}{{\cal H}}

\title{Bounds on the entropy generated  when  timing information\\
is extracted from microscopic systems}

\author{Dominik Janzing\thanks{Electronic address: janzing@ira.uka.de} 
   and 
Thomas Beth}
\address{Institut f\"ur Algorithmen und Kognitive Systeme, 
Universit\"at Karlsruhe, Am Fasanengarten 5,\\
    D--76\,131 Karlsruhe, Germany}

\begin{document}

\maketitle

\begin{abstract}We consider Hamiltonian quantum systems with energy bandwidth 
$\Delta E$ and show  that each measurement that  
determines the time up to an error $\Delta t$ generates at least 
the entropy $(\hbar/(\Delta t \Delta E ))^2/2$.
Our result
describes quantitatively to what extent all timing information
is quantum information in systems with limited energy. It 
provides a lower bound on the dissipated energy when
timing information of microscopic systems is converted to classical 
information. This is relevant for low power computation since it
shows the amount of  heat  generated whenever 
a band limited signal controls a classical bit switch.

Our result provides a general bound on the 
information-disturbance  trade-off for
von-Neumann measurements that  distinguish
states
on the orbits of  continuous unitary 
one-parameter groups with bounded spectrum.
In contrast, information gain without disturbance is possible for
some completely positive semi-groups.
This shows that readout of timing information can be possible without 
entropy generation if the autonomous  dynamical evolution
of the ``clock'' is dissipative itself.
\end{abstract}

\begin{multicols}{2}

\section{Timing information gain without disturbance?}

\label{In}
Listening to ``folklore versions'' of quantum mechanics one may
consider it as a key statement of quantum theory that there is no
measurement without disturbing the measured system.
However, the fact that  this is not true is well-known in 
modern quantum information theory and is, in some sense, the reason
why classical information exists at all although our world is quantum.
Consider a two-level system, i.e., a quantum system with Hilbert space $\C^2$
and denote its upper or lower state by $|1\rangle$ and $|0\rangle$, 
respectively. Assume that we know by prior information that the system
is not in a quantum superposition but only in one of the two states 
$| j\rangle$. Then the measurement with projections 
$P_0:=|0\rangle \langle 0|$ and $P_1:=|1\rangle \langle 1|$ 
show which state is present without disturbing it at all.
Here we have used the two-level system as {\it classical} bit.
The situation changes if the two-level system is used as {\it quantum} bit
(``qubit'') and is prepared in a quantum superposition 
$|\psi\rangle:=c_0|0\rangle+c_1|1\rangle$ where
the complex coefficients $c_0$ and $c_1$ with $|c_0|^2+|c_1|^2=1$ 
are unknown to the person who measures.
Then any von Neumann measurement
with projections $|\phi \rangle \langle \phi |$ and
$|\phi^\perp \rangle \langle \phi^\perp|$ will
on the one hand only provide some information about $|\psi\rangle$ 
and will on the other hand disturb the unknown state $|\psi\rangle$ since it  
``collapses'' to the state $|\psi\rangle $ with probability
$p:=|\langle \psi |\phi\rangle|^2$ and to the orthogonal state 
$|\phi^\perp\rangle$ with 
probability $q:=|\langle \psi |\phi^\perp\rangle |^2$.
From the point of view of the person who has prepared the state 
$|\psi\rangle$ and does not notice the measured result 
(``non-selective operation''), the measurement
process changes the density matrix of the system from
$|\psi\rangle \langle \psi |$ to 
\[
p |\phi \rangle \langle \phi | +q|\phi^\perp \rangle \langle \phi^\perp|\,,
\]
i.e., the measurement causes an entropy increase of
$\Delta S= -p\ln p - q\ln q$.

The general condition under which information about unknown quantum 
states can be gained without disturbing them is well-known
and reads as follows \cite{KI}:

Let $\rho$ be the unknown density matrix of a system. By prior
information one knows that $\rho$ is an element of a set
$\Gamma$ of possible states. Then one can get some information on $\rho$ if 
and only if there is a projection $Q$  commuting 
with all matrices in $\Gamma$ such that the value $tr( Q \rho )$ is not the
same for all $\rho \in \Gamma$.

As noted in \cite{viva2002}, this can never be the case if
the set $\Gamma$ is the orbit $(\rho_t)_{t\in \R}$ of 
a Hamiltonian system evolving according to
$\rho_t= \exp(-iHt) \rho \exp(iHt)$. 
This holds even if $\rho$ and $H$ act on an infinite dimensional Hilbert
space.
In this sense,
timing information is always to some extent quantum information
that cannot be read out without state disturbance.
It can only become classical information if either 
(1) prior information tells us that the time $t$ is an element
of some discrete set $\{t_1,t_2,\dots\}$ (see \cite{viva2002}) or (2) 
in the limit of infinite system energy  \cite{clock,viva2002}.

At first sight the statement that classical timing information 
can only exist in one of these two cases seems to be disproved by the
following dissipative ``quantum clock'':

Let $\rho_0:=|1\rangle \langle 1|$ be the upper state of a two-level 
system. Let the system's time evolution for positive $t$ 
 be described by the Bloch relaxation  (see e.g. \cite{ernst})
\[
\rho_t:= \exp(-\lambda t)|1\rangle \langle 1| + (1-\exp(-\lambda t))|0\rangle \langle 0|\,.
\]
Since all the states $\rho_t$ commute with the projections $P_0$ and $P_1$ 
one can certainly gain
some information about $t$ by the measurement with projections 
$(P_0,P_1)$.
However, this situation is actually the infinite energy limit since
semi-group dynamics of this form is generated by coupling 
the system to a heat bath 
of infinite size and infinite energy spectrum \cite{GardinerZoller}. 
Of course the fact that well-known derivations
of relaxation dynamics require heat baths with  infinite energy spectrum 
does not prove our
claim that this
is necessarily the case. This claim is rather an implication
of Theorem \ref{mainTh} in Section \ref{main}. 

The paper is organized as follows.
In Section \ref{main} we show quantitatively to what extent
information on the (non-discrete) time is always quantum
information as long as the clock is a system with limited energy bandwidth $\Delta E$.
Explicitly, we prove a lower bound 
on the entropy increase in the clock caused by von-Neumann measurements, i.e.,
measurements that are described by an orthogonal family of projections
where the state change of the system is described by the projection postulate.

Generalized measurement procedures 
are considered in Section \ref{Gen}.
They do not necessarily increase the entropy of
the measured clock since the process can include some kind of cooling 
mechanisms but
will lead to an increase of entropy of the total system which includes
the clock's environment. This has implications for
low power computation since it shows that the distribution of
timing information inherent in a microscopic clock
produces necessarily some phenomenological entropy.

In Section \ref{Tight} we discuss whether our bound is tight.
In Section \ref{Dec} we shall show that the result of Section \ref{Gen}
can be applied to the situation that a clock with limited energy controls
the switching process of a classical bit. Here  a classical bit is
understood as a two state quantum system on which decoherence takes
place on a time scale that is in the same order as the switching time
or on a  smaller scale. For low power computation, this 
proves which amount of dissipation is required
whenever the autonomous dynamics of a microscopic  device
controls a classical output.
In Section
\ref{Group} we apply the bound to other one-parameter groups.

\section{Entropy increase of a quantum 
clock caused by von-Neumann measurements}

\label{main}
For the moment we will restrict our attention to von-Neumann measurements.
For technical reasons we will assume them to have a finite set of 
possible outcomes. Hence the measurement is described by a family
$(P_j)$ of mutual orthogonal  projections on the system's Hilbert space
 $\cH$ that may be infinite dimensional.
The state of the system is described by a density matrix, i.e.,
a positive operator
with trace $1$ acting on $\cH$.
According to the projection 
postulate any state $\gamma$ changes to the post-measurement state
\[
\tilde{\gamma} :=\sum_j P_j \gamma P_j\,,
\]
if the {\it unselected} state is considered, i.e., the outcome
is ignored. 

The post-measurement state $\tilde{\gamma}$ 
coincides with $\gamma$ if and only if $\gamma$ commutes 
with all projections $P_j$.
Now we will compare the von-Neumann entropy of $\gamma$ and $\tilde{\gamma}$.
For any state $\gamma$ the von-Neumann entropy is defined as
\[
S(\gamma):=- tr(\gamma \ln \gamma)\,.
\]

The following Lemma shows that the measurement can never decrease the 
entropy:

\begin{Lemma}\label{KullLemma}
Let $\gamma$ be an arbitrary density matrix on a Hilbert space
and $(P_j)$ be a family of orthogonal projections with
$\oplus P_j=1$.
Set $\tilde{\gamma}:=\sum_j P_j \gamma P_j$
Then we have
\[
\Delta S: = S(\tilde{\gamma})- S(\gamma)=
K(\gamma || \tilde{\gamma})
\]
where $K(\gamma ||\tilde{\gamma})$ is the Kullback-Leibler distance
(or  Relative Entropy) between
$\tilde{\gamma}$ and $\gamma$. It is defined by \cite{OhyaPetz}
\begin{equation}\label{Kull}
K(\gamma ||\tilde{\gamma}):= tr(\gamma
\ln \gamma) - tr(\gamma \ln \tilde{\gamma} )\,.
\end{equation}
\end{Lemma}

\vspace{0.3cm}
\noindent
Proof: According to the definition of entropy
it is sufficient to show the equation $tr(\gamma \ln \tilde{\gamma})=
tr(\tilde{\gamma} \ln \tilde{\gamma})$.
The second term on the right in eq. (\ref{Kull}) is equal to
\begin{eqnarray*}
tr ( \sum_i P_i \gamma \ln (\sum_j P_j \gamma P_j)
=tr ( \sum_i P_i \gamma \ln (\sum_j P_j \gamma P_j) P_i)
\end{eqnarray*}
Note that 
$\ln (\sum_j P_j \gamma P_j)$ commutes with each $P_i$.
Hence we get
\begin{eqnarray*}
tr ( \sum_i P_i \gamma \ln (\sum_j P_j \gamma P_j) P_i) &=&\\
tr ( \sum_i P_i \gamma P_i \ln (\sum_j P_j \gamma P_j))&=&\\
tr(\tilde{\gamma} \ln \tilde{\gamma})
\end{eqnarray*}
This completes the proof. $\Box$
\vspace{0.3cm}

Since the entropy increase is the Kullback-Leibler distance between
the pre- and the post-measurement state we can obtain a lower bound
on $\Delta S$ in terms of the trace-norm distance between them:

\begin{Lemma}\label{Delta}
For the entropy increase $\Delta S$ we obtain the lower bound
\[
\Delta S\geq \frac{1}{2} \|\tilde{\gamma} -\gamma\|_1^2
\]
where
 $\|a\|_1:=tr( \sqrt{a^\dagger a})$ is the trace-norm of an arbitrary matrix 
$a$.
\end{Lemma}

\noindent
The proof is immediate using 
\[
K(\gamma|| \tilde{\gamma}) \geq 
\frac{\|\tilde{\gamma} -\gamma\|_1^2}{2}
\]
(see \cite{OhyaPetz}).

Now we consider the states $\rho_t$ on the orbit of the time
evolution and
show that the norm distance between
$\rho_t$ and $\tilde{\rho}_t$ is large at every moment where 
the outcome probabilities of the measurement $(P_j)$
change quickly, i.e., where the values $tr(i[H,P_j] \rho_t)$
are large. This is the essential statement  that is used to prove
the information-disturbance trade-off relation.
But first we have to introduce the energy bandwidth of a system.
For Hamiltonians with discrete eigenvalues it is just the difference
between the greatest and smallest eigenvalues of the
energy states with non-zero occupation probability. 
The generalization to  Hamiltonians with continuous and
discrete parts in the  spectrum reads:

\begin{Definition}
Let $H$ be the Hamiltonian of a quantum system and 
$(Q_E)_{E\in \R}$  be the spectral family corresponding to $H$, i.e.,
$Q_E$ projects onto the Hilbert space corresponding to energy values
not greater than $E$. Then $f(E):=tr(Q_E \rho_t)$ 
is the distribution function
of a  
(time-independent) probability measure $\mu$ which is called the spectral 
measure of $\rho_t$   corresponding to $H$.
Let $[E_{min},E_{max}]$ be the smallest interval supporting this spectral 
measure. Then $\Delta E:=E_{max}-E_{min}$ is the energy bandwidth
of the system in the state $\rho_t$.
\end{Definition}

By rescaling the Hamiltonian it is easy to see that
the time evolution of the state $\rho$ 
can equivalently be described 
by 
\[
H':=(Q_{E_{\max}}-Q_{E_{\min}}) H - \frac{1}{2}(E_{\max} -E_{\min}) {\bf 1}\,,
\]
since $\exp (-iH't) \rho \exp(iH't) =\exp (-iHt) \rho \exp (iHt)$.
Clearly, $\|H'\|\leq \Delta E /2$. 
The energy spread is decisive for our lower bound on the 
trace-norm distance between $\rho_t$ and $\tilde{\rho}_t$.

\begin{Lemma}\label{distanceGeschw}
Let $p_j(t):=tr(P_j \rho_t)$ be the probability of the measurement outcome
$j$ at time $t$. Let $\Delta E$ be the energy bandwidth of $\rho_t$.
Set 
\[
\|\dot{p}(t)\|_1:= \sum_j |\frac{d}{dt} p_j(t)|\,.
\]
Then we have 
\[
\| \rho_t -\tilde{\rho}_t  \|_1 \geq \frac{\|\dot{p}(t)\|_1}{\Delta E}\,.
\]
\end{Lemma}

\vspace{0.3cm}
\noindent
Proof:
For a specific moment  $t$ 
define the operator $R(t)$ by
\[
R(t):= s_j(t) P_j
\]
where $s_j(t)=1$ if $\dot{p}_j(t)\geq 0$ and $s_j(t)=-1$ if $\dot{p}_j(t)<0$.
Note that we have
\[
\|\dot{p}(t)\|_1 = \sum_j s_j(t) \dot{p}_j(t) =
tr(i[H,R(t)] \rho_t)\,.
\]
Furthermore easy computation shows that $tr( i[H,R(t)] \tilde{\rho}_t)=0$.
We conclude
\[
\|\dot{p}(t)\|_1 =tr(i[H,R(t)] (\rho_t-\tilde{\rho}_t))
\leq 2\|H\| \,\|\rho_t-\tilde{\rho}_t\|_1\,,
\]
since the operator norm of $R(t)$ is $1$.
We assume $\|H\|=\Delta E/2$ without loss of generality and have
\[
\|\rho_t-\tilde{\rho}_t\|_1 \geq  \frac{\|\dot{p}(t)\|_1}{\Delta E} \,.
\]
$\Box$
\vspace{0.3cm}

Now we are able to prove our main theorem.

\begin{Theorem}\label{mainTh}
Let the energy bandwidth $\Delta E$ of a quantum system be less than $\infty$.
Assume that the true time $t$ is in the interval $[0,T]$ for arbitrary $T>0$.
Let the prior probability for $t$ be the uniform distribution on $[0,T]$.
Let $(P_j)$ be the family of projections corresponding to a von-Neumann
measurement. Assume that the measurement has the time resolution
 $\Delta t$ in the following sense:
For each $t\in [0,T-\Delta t]$ there is a decision rule based on the
measurement outcome that decides whether the state $\rho_t$ 
or $\rho_{t+\Delta t}$  is present with error probability at most $1/4$.

Then the mean entropy increase  $\overline{ \Delta S}$ (averaged over the interval $[0,T]$)
caused by the measurement
is at least 
\[
\frac{1}{2}\Big(\frac{\hbar}{\Delta t \Delta E}\Big)^2\,.
\]
\end{Theorem}

Note that 
our definition of time resolution is not the usual one since
it does not require that the measurement distinguishes between
$\rho_t$ and $\rho_{t+2\Delta t}$, for instance.
However, it is exactly the definition of time resolution that we need
in the proof.

\vspace{0.3cm}
\noindent
Proof:
Each decision rule distinguishing between $\rho_t$ and
$\rho_{t+\Delta t}$ that is based on the measurement outcome
is of the following form: 
If the outcome is $j$ it decides for $t$ with probability $q_j$.
If the true state is $\rho_t$ it decides for $t$ with probability
$
\sum_j q_j p_j(t)$ and if the true state is $\rho_{t+\Delta t}$ it decides
erroneously for  $t$ with probability 
\[
\sum_j q_j p_j (t+\Delta t) =\sum_j q_j p_j(t) +
\sum_j q_j \int_t^{t+\Delta t} \dot{p}_j(t') dt'\,.
\]
The difference between the probability to decide correctly for $t$ and to
decide erroneously for $t$ is at least $1/2$ by assumption.
Hence 
\[
|\int_t^{t+\Delta t} \sum_j  q_j p_j(t') dt'| \geq 1/2\,.
\]
Note that $2q_j-1$ is in the interval $[-1,1]$ for all $j$. Therefore
\[
|\sum_j (2q_j -1) \dot{p}_j(t)| \leq  \sum_j |\dot{p}_j(t)| \,.
\]
Obviously we have $\sum_j \dot{p}_j(t)=0$ since $\sum_j p_j(t)=1$.
We conclude
\[
\|\dot{p}(t)\|_1 \geq 2 |\sum_j q_j \dot{p}_j(t)|\,,
\]
hence
\[
\int_t^{t+\Delta t} \|\dot{p}(t')\|_1 dt' \geq 1 \,.
\]
Therefore we find that the average of $\|\dot{p}(t')\|_1$ over the interval
$[t,\Delta t]$
is at least $1/\Delta t$.
Since this holds for every $t\in [0,T-\Delta t]$ the average of
$ \|\dot{p}(t)\|_1$ over the whole interval $[0,T]$ is at least
$1/\Delta t$.
We denote this average by
\[
\overline{\|\dot{p}(t)\|_1}\,.
\]

For the average entropy generation we have
\[
\overline{ \Delta S} =\frac{1}{T}\int_0^T S(\tilde{\rho}_t) -S(\rho_t)\,
dt \geq \frac{1}{2} \overline{\|\tilde{\rho}_t -\rho_t\|_1^2} 
\]
by Lemma \ref{Delta}. Note that the average of the square
is at least  the square of the average.
We conclude
\[
\overline{\Delta S} \geq \frac{1}{2} \overline{\|\tilde{\rho}_t -\rho_t\|_1}^2\,.
\]
Due to Lemma \ref{distanceGeschw}
we have
\[
\overline{\Delta S} 
\geq \frac{1}{2} \frac{1}{(\Delta t \Delta E)^2}\,,
\]
as long as we measure the energy spread in units of $\hbar$.
In physics, it is more common to use SI-units which lead to 
\[
\overline{\Delta S} 
\geq \frac{1}{2} \Big(\frac{\hbar}{(\Delta t \Delta E}\Big)^2\,.
\]
$\Box$
\vspace{0.3cm}

\section{Entropy generated by generalized measurements}

\label{Gen}
The assumption that every measurement is described by projections
$(P_j)$ and that the effect of the measurement is given by
the projection postulate
\[
\rho \mapsto \sum_j P_j \rho P_j
\]
is not general enough.
In general, a measurement 
can be described by a positive operator valued measure (POVM)
(see \cite{Da76} for
the most general description of measurements).
If the set of possible outcomes is finite, a POVM
is defined by a family $(M_j)$ of positive operators
with $\sum_j M_j={\bf 1}$
such that $tr(M_j \rho)$ is the probability for the outcome
$j$. Furthermore the connection between the POVM and the effect on the state
is given by the following consistency condition:
There exist completely positive maps $G_j$ such that
$tr(G_j(\rho))=tr(M_j \rho)$ and  that, given the outcome $j$, the 
post-measurement state is $G_j(\rho) /tr(M_j \rho)$.
The unselected post-measurement state is hence  given by
\[
\tilde{\rho}:= \sum_j G_j(\rho)\,.
\] 
Clearly, in this general setting it is even possible
that the measurement can decrease the entropy of the clock since
the maps $G_j$ may include processes that are cooling mechanisms 
for the clock
and transport entropy to the environment. 
      
However, we will show that each time measurement leads unavoidably to
an entropy increase in the total system consisting of the 
measurement apparatus
and the clock. 
Let $\cH_m$ be the Hilbert space of the measurement apparatus and 
be $\hat{H}$ be
the Hamiltonian  
its  free time evolution as long as no  measurement interaction is active.
We assume the state  $\gamma$ of the apparatus 
to be invariant under its free evolution. 
Otherwise the 
measurement apparatus would be a clock in its own since it  contained
some information about the time $t$.
Then we switch on the measurement interaction which leads to
a unitary transformation $u$  on the space 
$\cH \otimes\cH_m$  
(the definition of the apparatus includes its environment such that
the total evolution is unitary).
This process is often called a {\it pre-measurement}.
The different outcomes $j$ correspond to mutual orthogonal subspaces
of $\cH_m$ (``pointer states'').
Let $(Q_j)$ be the projections onto these subspaces.
Now we assume that decoherence on the measurement apparatus' pointer takes
place \cite{Zurek1,Guilini,Zurek2}. The process given by the unitary 
pre-measurement process $u$
and the decoherence is described by
\[
\rho\otimes \gamma  \mapsto  
\sum_j ({\bf 1}\otimes Q_j) u (\rho \otimes \gamma) u^\dagger   
({\bf 1}\otimes Q_j) 
\]
The entropy of this state is clearly the same as the entropy of
\[
\sum_j u^\dagger({\bf 1}\otimes Q_j) u 
(\rho \otimes \gamma) u^\dagger   ({\bf 1}\otimes Q_j) u \,.
\]
By defining $P_j := u^\dagger ({\bf 1} \otimes
 Q_j) u$ we have reduced the problem
to the problem in Section \ref{main} with the only generalization
that  $(P_j)$ is not a family of projections acting on the Hilbert space 
$\cH$ 
of the clock
but on the Hilbert space of clock + measurement 
apparatus where the apparatus is at rest.
Now we show that the stationarity of  the apparatus state 
ensures that the bound in Theorem \ref{mainTh} holds for 
the considered  kind of generalized measurement
as well. 

Formally, we claim:

\begin{Theorem}\label{mainThGen}
Let $\rho\otimes \gamma$ be a density matrix on a Hilbert space 
$\cH \otimes \hat{\cH}$. Let the time evolution 
of the composed system be given by the Hamiltonian
$H_c:=H\otimes {\bf 1} + {\bf 1} \otimes \hat{H}$ with $[\gamma,\hat{H}]=0$.
Let $(P_j)$ be a von-Neumann measurement 
acting on the Hilbert space $\cH \otimes \hat{\cH}$.
Let $\Delta E$ be the energy bandwidth of the left component of the 
composed system (given by the state $\rho$ with Hamiltonian $H$).
Let the measurement have the  time resolution $\Delta t$ 
in the sense of Theorem \ref{mainTh}. Then we have the following lower bound
on  the average 
entropy increase $\overline{\Delta S}$ of the composed system
caused by the measurement:
\[
\overline{\Delta S} \geq \frac{1}{2}\Big(\frac{\hbar}{\Delta t \Delta E}\Big)^2\,.
\]
\end{Theorem}
\vspace{0.3cm}

\noindent
Proof:
Clearly, the Hamiltonian $\hat{H}$ is irrelevant 
for the  evolution of the state $\rho\otimes\gamma$ due to
\[
\exp(-iH_ct) (\rho \otimes \gamma) \exp(iH_c t)=
\exp(-iHt) \rho \exp(iHt) \otimes \gamma\,.
\]
Hence we can treat the evolution as if it was implemented by
the Hamiltonian $H\otimes {\bf 1}$, which can assumed to be bounded
as in Section \ref{main}.
$\Box$.
\vspace{0.3cm}

Note that our arguments do not require that the 
time evolution $u$ takes place
on a smaller time scale than $\Delta t$.
The reason is that we have argued that $P_j:=u^\dagger 
({\bf 1}\otimes Q_j) u$ 
may formally be considered
as projections of a  
measurement performed {\it before} the interaction was switched on.

Our results in Section \ref{main} and \ref{Gen}  may be given an additional 
interpretation as information-disturbance trade-off relation.
Lemma \ref{KullLemma} shows that the entropy increase caused by the measurement
is the Kullback-Leibler distance between pre- and post-measurement state.
Information-disturbance trade-off relations are an important part of
quantum information theory \cite{Fu96b}.  Unfortunately, 
Theorem \ref{mainTh}
is restricted to von-Neumann measurements. Theorem \ref{mainThGen}
extends the statement to general measurements as far as
the disturbance of the total state $\rho_t \otimes\gamma$
of the ``clock'' and the ancilla system 
is considered. However, this is interesting from the thermodynamical
point of view  taken in this 
article but not in the setting of information-disturbance
trade-off. In the latter setting, the disturbance of the ancilla state
that is used to implement a non-von-Neumann measurement is not of interest.
Therfore we admit, that our bound on the state disturbance
does only hold for von-Neumann measurements.

\section{How tight is the bound?}

\label{Tight}

The entropy increase  predicted by our results can never
be greater than $1/2$.
This can be seen by the following argument. Using the 
definitions of the proof of Lemma \ref{distanceGeschw} we have
\begin{eqnarray*}
\|\dot{p}(t)\|_1 &=& tr( i[H,R(t)]\rho_t) \leq
2\|H\|\,\|R(t)\| =\Delta E \|R(t)\|\\&=& \Delta E\,.
\end{eqnarray*}

Note also the connection to 
the Heisenberg uncertainty principle
\[
\Delta' E \Delta' t \geq \hbar /2\,.
\]
where $\Delta' E$ is the energy spread of the system,  i.e., the standard
deviation of the energy values and $\Delta' t$ the standard deviation
of the time estimation \cite{BCM96} based on any measurement.
However, using the symbols $\Delta'$ instead 
of $\Delta$, we emphasize that these definitions do not agree with ours.
Note that the energy spread $\Delta'E$ is {\it at most} the energy bandwidth
$\Delta E$. But $\Delta' t$ can exceed
 our time resolution
$\Delta t$ by an arbitrary large value. 
This shows the following example:
assume the clock to be a two-level system with period
$\tilde{T}$. 
Set $\rho_t:=|\psi_t\rangle \langle \psi_t|$ with
\[ 
|\psi_t\rangle:= \frac{1}{\sqrt{2}}(|0\rangle 
+ \exp(i2\pi/\tilde{T})|1\rangle)\,.
\]
Given the prior information that
the time is in an interval $[0,T]$ with $T\gg \tilde{T}$  we can only
estimate the time up to multiples of $\tilde{T}$
and obtain an error in the order $T$.
In contrast, the
time resolution $\Delta t$ in our sense is $\tilde{T}/2$ since we can 
certainly distinguish between the state at the time $t$ and at $t+\tilde{T}/2$.
However, we conjecture that similar bounds as in Theorem \ref{mainTh} 
on the entropy generation
can be given for systems where the energy spectrum is {\it essentially} 
supported by an interval $[E_{\min},E_{\max}]$.
Therefore our results suggest the  following 
interpretation: every measurement that allows to estimate the time
up to an error that is not far away from the Heisenberg limit 
produces a non-negligible amount of entropy.
The following example suggests that the necessary entropy generation
may even be much above our lower bound if ``time measurements''
are used that are extremely close to the Heisenberg limit.
Consider the wave function of a free Schr\"{o}dinger particle moving on 
the real line. A natural way to measure the time would be
to measure its position. We may use, for instance, 
von-Neumann measurements that
correspond to a partition of the real line into intervals of length
$\Delta x$. Assume one wants to improve the time  accuracy by
decreasing $\Delta x$ arbitrarily. 
The 
advantage for the time estimation  
is small if $\Delta x$ is smaller than the actual 
position uncertainty of the particle.
However, if the state is  pure it is easy to see that 
the generated entropy  
goes to infinity for $\Delta x\to 0$.

The following example shows that the entropy generation in a time measurement
can really go to zero when $\Delta E$ goes to infinity.
Consider the Hilbert space $\cH:=l^2(\Z)$ 
of square sumable functions over the set of integers.
Let $|j\rangle$ with $j \in \Z$ be the canonical basis vectors.
Let the Hamiltonian be the diagonal operator
\[
H |j \rangle := j |j \rangle\,.
\]
Consider the state 
\[
|\psi\rangle := \frac{1}{\sqrt{k}} \sum_{j=0}^{k-1} |j\rangle\,.
\]
By Fourier transformation $\cH$ is isomorphic to the set $L^2(\Gamma)$ of
square integrable functions on the unit circle $\Gamma$ 
and the dynamical evolution is the cyclic
shift on $\Gamma$. The period of the dynamics is $T=2\pi$. 
In this picture, $|\psi\rangle$ is a wave package 
that has its maximum at the angle $\phi_t=t$
with an uncertainty $\Delta \phi$
in the order  $1/k$.
We assume to know that the true time is in the interval
$[0,2\pi)$
and 
construct a measurement that allows to decide 
for each $t \in [0,\pi)$ whether
the true time is $t$ or $t +\pi$ with confidence that is increasing
with $k$.
We use a measurement with projections $P_1,P_2,P_3,P_4$ 
projecting on the four sectors of the unit circle.
If the angle uncertainty is considerably smaller than $\pi/2$
this measurement can clearly distinguish between 
$t$ and  $t + \pi$ with high confidence.
Hence, for $k$ large enough, 
we get $\Delta t=\pi$ as time resolution of the measurement.
Note that we have only non-negligible entropy generation at that
moments where the main part of the wave packet crosses the border between
the sectors, i.e., when there are two sectors containing  
a non-negligible part of 
the wave function. If, for instance, each sector contains about
one half of the probability, we generate the entropy one bit, i.e.,
the entropy $\Delta S=\ln 2$ in natural units.

The probability that the measurement is performed 
at a time  in which non-negligible entropy generation takes place
is about $2\Delta \phi /\pi$. The average  entropy generation decreases
therefore for increasing $k$, i.e., for increasing $\Delta E=k-1$.
Let $p_j^k(t)$ for $j=1,\dots,4$ be the probabilities for the $4$ possible 
outcomes when the wave packet with energy spread $k-1$ is measured at the 
time instant $t$.
There are four moments where the maximum of the wave packet is exactly 
on the
border of the sectors. The probability  to meet these
times is zero. For all the other times $t$ there is one $j$ such
that $1-p_j^k(t)$ tends to zero with $O(1/k^2)$ by standard Fourier analysis
arguments. Note that a measurement at time $t$ produces the entropy
\[
\Delta S= -\sum_j  p_j^k(t)\ln p_j^k(t)\,.
\]
We conclude by elementary analysis that 
$\Delta S$ tends to zero with $O((\ln k)/k^2)$ for all times $t$ 
except from $4$ irrelevant values $t$ 
and therefore the average entropy generation
tends to zero with $O((\ln k)/k^2)$.
Note that the decrease of  the lower bound of Theorem 
\ref{mainTh} is asymptotically 
a little bit faster since it is
$O(1/k^2)$ (due to $\Delta E=k-1$).

\section{Controlling a classical 
bit switch by a
microscopic clock}

\label{Dec}
Now we consider the system consisting of the clock, its environment
and the classical bit. For the moment we ignore the fact 
that the bit is quantum and claim that the two logical states $0$ and $1$ 
correspond to an orthogonal decomposition
\[
\cH=\cH_0 \oplus \cH_1
\]
of subspaces of the Hilbert space of the  composed system.
If $\cH_0$ and $\cH_1$ are non-isomorphic as Hilbert spaces 
we extent the smaller space such that they are isomorphic. 
Then we can assume without loss of generality
$\cH$ to be of the form $\cH_0\otimes \C^2$ without assuming that 
the bit is  physically realized by a two-level quantum system.
The fact that the bit is classical will now be described by the fact
that it is subjected to decoherence, i.e., that
all superpositions between $|0\rangle$ and $|1\rangle$ are destroyed and
changed to mixtures on a time scale that is not larger 
than the switching time.
Decoherence keeps the diagonal values of the density matrix
whereas its non-diagonal entries decay.
If the process is a uniform time evolution, i.e.,
given by a semi-group dynamics,
decoherence of the two-level system is given by an exponential
decay of both non-diagonal entries:
\begin{eqnarray*}
\gamma_t&:=&\gamma_{00}|0\rangle \langle 0| +
\exp(-\lambda t)\gamma_{01}|0\rangle \langle 1| \\&+&
\exp(-\lambda t)\gamma_{10}|1\rangle \langle 0| +
\gamma_{11}|1\rangle \langle 1| \,,
\end{eqnarray*}
where $\gamma_{ij}$ are the coefficients of the density matrix.
Note that it is assumed that the effect of the environment 
does not cause any bit-flips in the system but only destroys 
coherence. The parameter $\lambda\geq 0$ defines the decoherence rate.
Let $P_0$ and $P_1$  be the projections onto the states
$|0\rangle $ and $|1\rangle$, respectively.
The decoherence process can be simulated by measurement processes
that are performed at randomly chosen time instants.
This analogy is explicitly given as follows:
Let 
\[
\tilde{G}(\gamma):=P_0 \gamma P_0 + P_1 \gamma P_1
\]
be the effect of the measurement that distinguishes the two logical states.
Then we have
\[
\gamma_t = \lim_{n\to \infty} ((1-\lambda/n) id+ \lambda  \tilde{G}/n)^n 
(\gamma)
=\exp(\lambda t (\tilde{G}-id)) (\gamma)\,.
\]
The second expression provides the following intuitive approximation
of the process: In each small time interval of length $1/n$ a measurement
is performed with probability $\lambda/n$.
Let $G:={\bf 1}\otimes \tilde{G}$ be the extension
of $\tilde{G}$ to the total system.
We assume that the dynamical evolution of the total system
is generated by
\begin{equation}\label{Fdef}
F:= i[.,H]+  (G-id)\,,
\end{equation}
i.e., the decoherence of the bit is the only contact of the system
to its environment.
Define the switching time as the length of the time interval 
$[0,\Delta t]$
where the probability of one of the logical states 
changes from $1/4$ to $3/4$.
The reason that we do not assume it to switch between $0$ and $1$ is that
this is impossible within a finite time interval with a Hamiltonian
that has limited spectrum. This can be seen by the fact that
all expectation values with Hamiltonians
of lower (!) bounded spectrum are analytical functions \cite{Hegerfeldt}.
Consider the case that the decoherence time is small compared to
the switching time $\Delta t$, i.e., $\lambda \Delta t \ll 1$.
Then the probability to have  more than  one measurement 
during the switching process is small (of second order in $\lambda$) 
and the probability that
one measurement occurs is about $\lambda \Delta t$.
Given that it occurred, the probabilities of 
its time $s$ of occurrence is equally distributed in $[0,\Delta t]$.
Since entropy is convex the entropy generated by the switching process
is at least  $\lambda \Delta t$ times the entropy that is generated
if a measurement has occurred during the switching process.
Also by convexity arguments, we conclude that the entropy
generated by the process ``perform a measurement at a randomly chosen 
(unknown) time
instant in $[0,\Delta t]$'' is at least the average entropy generation
if the time instant is known.
The latter situation meets exactly the assumptions of Theorem
\ref{mainTh} with the following parameters:
Set $T:=\Delta t$, i.e., 
assume that the prior information is that the true  time is in
$[0,\Delta t)$. Furthermore the switching time $\Delta t$ coincides  with
the time resolution $\Delta t$ of Theorem \ref{mainTh} since
reading out the logical state can distinguish between
the times $0$ and $\Delta t$ with error probability at most $1/4$.
Taking into account that a measurement occurs only
with probability $\lambda \Delta t$ we multiply the bound
of Theorem \ref{mainTh} by this factor and obtain:

\begin{equation}\label{coh}
\Delta S\geq \frac{\lambda \Delta t}{2} \Big(\frac{\hbar}{\Delta t \Delta E}
\Big)^2 =\frac{\lambda \hbar^2}{2 \Delta t (\Delta E)^2} \,.
\end{equation}

Note that we do not speak about {\it average} entropy generation
since we consider (in contrast to Theorem \ref{mainTh}) 
the entropy  of a single density matrix which is already an average
over a set of  density matrices
which are obtained when the measurement occurs at different times. 

Consider now the case that the decoherence time is so small that
more than one measurement during the switching process is likely.
Then we have to worry about the post-measurement state and whether
its energy spectrum is still bounded. If not, we cannot apply
our arguments. If it is, we can use the bound (\ref{coh}) nevertheless.
This is less obvious than it may seem at first sight.
The total dynamics is assumed to be the semi-group evolution
\[
\rho_t:=\exp(Ft)(\rho)
\]
with $F$ as in eq. (\ref{Fdef}).
Now we investigate the amount of entropy generated by a measurement
at time $t$. Note that we do not have a lower bound by Theorem \ref{mainTh}
since the dynamical evolution leading to the state $\rho_t$ was
not Hamiltonian. However,  the bound
\[
\Delta S \geq \Big(\frac{\|\dot{p}(t)\|_1}{\Delta E}\Big)^2
\]
holds nevertheless. This is seen by the observation
that the generator $G$ is irrelevant 
if the
proof of Lemma \ref{distanceGeschw} should be converted
to the situation here.
With $\tilde{\rho}_t:=\sum_j P_j \rho_t P_j$
we have 
\[
\dot{p}_j (t) =tr(P_j F(\rho_t -\tilde{\rho}_t))=tr(P_j i[\rho_t
-\tilde{\rho}_t,H] )
\]
due to $tr(P_j (G-id) \gamma)=0$ for every state $\gamma$.
It should be emphasized that this argument would fail if the two-level system
was not only subjected to decoherence but also to relaxation.
If $L$ is the generator of a relaxation process
that causes directly transitions from
the state $|1\rangle$ to the state $|0\rangle$ 
the equation $tr( P_j L \gamma)=0$ does not hold, i.e.,
$L$ would not be irrelevant for our proof.
This is consistent with the observation in Section \ref{In} 
that the ``relaxation clock'' can be read out without 
generating entropy {\it by the measurement}.
We conclude that the bound (\ref{coh}) holds for arbitrarily high decoherence
rate. Formally, the bound predicts infinite entropy production
if $\Delta t$ and $\Delta E$ are constant and $\lambda \to \infty$.
However, this asymptotics is meaningless due to the quantum zeno effect
(see e.g. \cite{Gurvitz}).
Infinite decoherence would stop the switching process completely.
Consider a short time $t$ after a measurement has been performed
on the state. Then the derivative of the probabilities $p_j$ 
corresponding to a second measurement is only of the order 
$\Delta E t$. The essential consequence is that 
in particular for small $\Delta E$ when our bound would predict
large entropy generation fast switching processes are impossible.

In order to show how to apply our bound 
to realistic situations
consider the following  example:
A light signal (guided by an optical fiber, for instance)
is sent to an apparatus that contains a two-level system.
The incoming signal 
triggers the transition $|0\rangle \mapsto |1\rangle$ 
from the lower to the upper level (see Fig.1).

\begin{figure}
\epsfbox[0 -10 202 73]{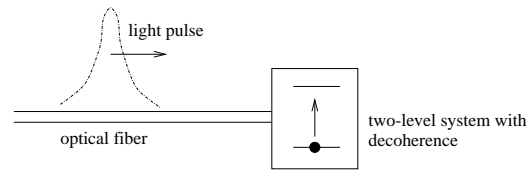}
\caption{Classical bit switch controlled by a light pulse.
For time $t\to-\infty$ the time evolution is approximatively 
the free evolution of the light field without any interaction with the 
apparatus.}
\end{figure}

The apparatus may be a huge physical system
and we are only interested in the fact that the {\it transition time}
is given by the  incoming light signal, the energy may be provided
by the apparatus itself.
Assume the light signal 
consists of photons with frequency bandwidth 
$\Delta \omega$. The signal is assumed to  contain at most $k$ photons.
Than the energy bandwidth of the signal is at most 
$\Delta \omega k$.  
For $t\to-\infty$ the time evolution of the total system
is evolving approximatively as
\[
\rho_t \otimes\gamma
\]
where $\gamma$ is the (stationary) state of the apparatus.
When the signal meets the apparatus the time evolution is 
given by an unknown Hamiltonian $H_c$ of the total system
combined with the decoherence of the two-level system.
The energy bandwidth of the total system according to $H_c$ is unknown.
The ``measurement'' on the two-level system takes definitely place 
at a moment where the total evolution of the system is generated
by the unknown Hamiltonian $H_c$.
Nevertheless we can apply the bound of Theorem \ref{mainThGen}.
In Section   \ref{Gen} we have argued that Theorem \ref{mainThGen} 
does not assume that the measurement interaction was only switched on 
during a time interval that is small compared to the time resolution
$\Delta t$. We observed that the measurement can formally be 
considered as a measurement that had been performed {\it before} 
the interaction had been switched on.
Analogously, we argue that the unknown interaction 
between light field and apparatus is irrelevant:
Let $\sigma_t$ be the state of the total system at the time $t$.
Let the switching process happen during the interval 
$[0,\Delta t]$. Let $(u_t)_{t\in \R}$ 
be the time evolution of the total system implemented by a possibly unbounded
Hamiltonian.
The entropy 
generated by ``measurements'' $P_j$  performed 
on the state $\sigma_t$ during this period
due to the decoherence is the same as 
produced by $u^\dagger_s P_j u_s$ performed on the state
$u_s\sigma_t u_s^\dagger$. For $s\to -\infty$ the family of states
$(u_s \sigma_t u_s^\dagger)_{t\in [0,\Delta t]}$ evolve
approximatively like the free evolution
$
\rho_t \otimes \gamma\,
$
of the light field. 
We conclude that the bound in eq. (\ref{coh})
can be applied and the relevant energy bandwidth 
is the energy bandwith of the free light field.
The entropy generation is at least
\[
\frac{\lambda}{2\Delta t (n \Delta \omega)^2}\,.
\]

Statements of this kind may be relevant in future computer technology
when miniaturization reduces signal energies on the one hand
and requires on the other hand
reduction of power consumption. Our results show that
reduction of signal energy down to the limit of 
the energy-time uncertainty principle leads unavoidably to  
heat generation as long as the signal control
{\it classical} bits.
The entropy generation $\Delta S$ leads to an energy loss
of $\Delta S k T$ (where  $k$ is Boltzmann's constant and $T$ is 
the absolute temperature) due to the second law of thermodynamics.

\section{Generalization to other one-parameter groups}

\label{Group}
Obviously, our results generalize to other one-parameter groups since
the proofs do not rely on the interpretation of the unitary group
$\exp(-iHt)$ as the system's {\it time} evolution.

Consider for instance the case that the 
momentum of a Schr\"{o}dinger wave package
is restricted to the interval $[p, p+\Delta p]$.
Let $\rho$ be the particle's density matrix and $\rho_x$ the state
translated by $x\in \R$.
Then we conclude that every measurement that is suitable to 
distinguish between $\rho_x$ and $\rho_{x+\Delta x}$ 
in the sense of Theorem \ref{mainTh} 
produces at least the entropy
$(\hbar/(\Delta x \Delta p))^2/2$. 
Note that the measurement is not necessarily
a position measurement, it has not even to be compatible 
with the position operator.
In this sense, our results may be interpreted as a kind of 
``generalized 
uncertainty relation''.

Another very natural application is to consider the group of rotations
around a specific axis. Consider a  spin-k/2 particle and
the group of rotations
\[
(\exp( iL_z\alpha/\hbar))_{\alpha\in \R}
\]
on its Hilbert space $\C^{k+1}$, where $L_z$ is the operator of
its angular momentum in $z$-direction. We have $\Delta L_z=k \hbar$
and conclude the following:
Each measurement that  distinguishes between $\rho_\alpha$ and
$\rho_{\alpha + \Delta \alpha}$ with error probability at most $1/4$
produces at least the entropy
$1/(2(k \Delta \alpha)^2)$.

\section{Conclusions}

Our bound on the entropy that is generated when information
about the actual time is extracted from quantum system holds only
for Hamiltonian time evolution. A simple counterexample in Section \ref{In} 
has shown that time readout without state disturbance is possible for some
dissipative semi-group dynamics.
This leads to an interesting question:
In physical systems, each dissipative semi-group dynamics that is induced
by weak coupling to a reservoir in thermal equilibrium
is unavoidably accompanied by some loss of free energy.
This shows that the loss of free energy caused by the time measurement
can only be avoided by systems that loose energy during its
autonomous evolution. It would be desirable to know whether there is
a general lower bound (including dissipative dynamics)
on the total amount of free energy 
that is lost as soon as timing information is converted 
to classical information.

\section*{Acknowledgments}

Thanks to Thomas Decker for helpful comments.
This work has been supported by grants of the DFG
project ``Komplexit\"{a}t und Energie'' of the ``Schwerpunktprogramm
verlustmarme Informationsverarbeitung'' Be 887/12.

\end{multicols}


\end{document}